# Automatic dual air kerma strength treatment planning for focal low-dose-rate prostate brachytherapy boost using dosimetric and geometric constraints

S. Sara Mahdavi[1], Michael D. Peacock[1], William J. Morris[1], Ingrid T. Spadinger[2]

[1] Department of Radiation Oncology, BC Cancer – Vancouver Centre, Vancouver, BC, Canada

[2] Department of Medical Physics, BC Cancer – Vancouver Centre, Vancouver, BC, Canada

Abstract

**Purpose:** Automatic planning for focal low-dose-rate prostate brachytherapy (LDR-PB) boost.

**Methods:** A simulated annealing approach was utilized for creating automatic focal LDR-PB boost plans using both geometric (i.e. adherence of isodose contours to structure contours) and dosimetric constraints. In 46 patients, four unilateral focal lesion scenarios were considered as the boost region, in the anterior/posterior and superior/inferior quadrants. Plans consisted of seeds with two air kerma strengths (0.740U and 0.351 or 0.354U) to achieve required dose coverage on these smaller focal targets. Plan characteristics (including PTV $V_{100}$ and $V_{150}$, Rectum $V_{50}$, Urethra $V_{150}$, and number of needles and seeds), were compared in terms of type of needle loading (dual vs. single air kerma needles), boost quadrant and prostate volume. Plans were blindly ranked on 20 consecutive cases by two experienced radiation oncologists.

**Results:** Differences in plans generated using the two needle loading techniques were not clinically significant, particularly in terms of the PTV $V_{100}$ and $V_{150}$ and Urethra $V_{150}$, with single air kerma loaded needles potentially being more practical. A higher rectal dose was observed in the posterior boost quadrants ($V_{50}$ >0.18), particularly in larger prostates ($\geq$40 cm$^3$). In smaller prostates (<25 cm$^3$), obtaining adequate dose coverage while sparing the urethra was more challenging. 88% of the plans were considered clinically acceptable without (46%) or with (mostly minor) modifications.

**Conclusions:** The novel approach of using dual air kerma seeds combined with geometric constraints within a simulated annealing-based automatic planning framework can lead to focal LDR-PB plans with high acceptability.

## INTRODUCTION

Low-dose-rate prostate brachytherapy (LDR-PB) is an effective method for prostate cancer (PCa) treatment, both as monotherapy for low to intermediate-risk PCa and as combined treatment in intermediate to high-risk PCa [1, 2]. The randomized controlled trial ASCENDE-RT [1] performed at BC Cancer showed unprecedented levels of biochemical control for men with intermediate and high-risk PCa treated with combined EBRT and LDR-PB boost. However, these benchmarks were accompanied by a higher incidence of adverse genitourinary side effects [3].

The OPTiMAL clinical trial (ClinicalTrials.gov: NCT03836196), open to accrual at BC Cancer – Vancouver, aims at reducing the side effects observed in ASCENDE-RT. In ASCENDE-RT the LDR-PB boost component delivered a minimum peripheral dose (mpd) to the entire gland and 150% of the mpd to the entire peripheral gland. In fact, treatment of the entire gland is common in both LDR-PB monotherapy and combined treatment due to limitations in tumor localization. As a result of eradicating healthy tissue as well as the tumor, particularly in the rectum and urethra, treatment side effects are common. In OPTiMAL mapping biopsy is integrated with multi-modality imaging techniques for target localization and to achieve minimal morbidity with cancer control, 'focal' LDR-PB boost is applied. That is, for selected patients with localized PCa (ideally) only the diseased region receives a contiguous high dose boost to 150% of the mpd.

The need for a higher dose (boost) to a generally small volume can be achieved either by higher seed density or the use of high air kerma strength sources (higher than 0.4 U, a typical value used in our centre). In [4, 5] we showed that a dual activity planning approach for focal treatment is possibly the practical solution for limiting the density of planned sources and the proximity of high strength sources to organs at risk, while delivering the prescribed dose to the generally small PTV.

Another challenge in LDR-PB in general is the manual approach to planning. Automatic planning can provide a fast and reliable solution for both standard pre-planning and intra-operative pre-planning [6]. Despite early developments in automatic brachytherapy treatment planning for PCa, the focus appears to be more on high-dose-rate (HDR) brachytherapy [7, 8] rather than LDR-PB. Most proposed methods for LDR-PB are aimed at treating the entire gland [9-12] and with the literature on focal LDR-PB optimization being scarce [13].

In this paper we propose a first automatic method for focal LDR-PB boost planning for treatment of high-risk PCa using dual activity seeds and 'geometric constraints'. Standard planning guidelines applied at our centre are used for initialization and optimization of this simulated annealing-based algorithm. However, satisfactory dose constraints do not guarantee 'acceptable' and implantable plans from a physician's point of view (e.g. low needle density to avoid excessive trauma, and smoothness of the isodose surfaces). Hence, we have also included 'geometric' planning constraints to produce more acceptable plans.

**METHODS**

Inspired by Pouliot et al. [11] simulated annealing (SA) [14] is used for plan optimization. The goal of the optimization algorithm is to find the most suitable brachytherapy seed positions resulting in a treatment plan that meets the required clinical dosimetric and geometric criteria.

**Data set and Contour definitions**

Institutional ethics approval was acquired for use of data in this study. Trans-rectal ultrasound (TRUS) image sets with a variety of prostate shapes and sizes were chosen from our larger pool of standard whole-gland LDR-PB patient database (volume range: 11.4- 56.2 cm$^3$, median: 37.0 cm$^3$). TRUS images contain the prostate gland and one slice superior and inferior to the gland with 0.5 cm slice spacing. The planning target volume (PTV) is the clinical target volume (CTV- the prostate and a portion of the seminal vesicles) plus a margin of 0.35 cm anteriorly and laterally, 0.5cm superiorly and inferiorly, and no margin posteriorly (Figure 1). The rectum is contoured, and the urethral path is estimated according to Bucci et al [15].

The BK Medical (Peabody, MA, USA) 5mm × 5mm template, is overlaid on the TRUS images and is used as the seed location points within the plan (Figure 1).

We consider quadrant based focal LDR-PB boost with the targeted lesion located unilaterally in one of the four quadrants: anterior-inferior (AIQ), anterior-superior (ASQ), posterior-inferior (PIQ), and posterior-superior (PSQ). Quadrants are defined as approximately a quarter of the transverse PTV minus a portion of the gland midline and urethra [5]. The superior-inferior length of each quadrant is 2/3 of the length of the PTV from either end. Assuming lateral symmetry in anatomy, we consider only lesions located on the right side of the gland (Figure 1). The use of quadrants here is to maintain consistency in analysis among cases. This approach can be used on boost regions of other shapes and sizes obtained by means such as MRI or biopsy.

The PTV is thus divided into "PTV$_H$": one of the above four quadrants to be irradiated with the higher 150% mpd, and "PTV$_L$": the remainder of the PTV receiving 100% of the nominal prescribed dose (110 Gy here).

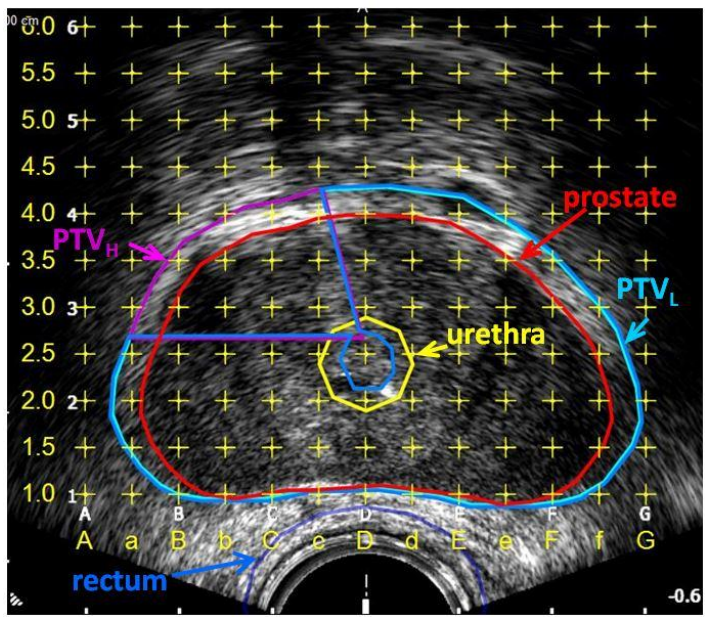

Figure 1: A transverse TRUS image of the prostate mid-gland, showing the template, the prostate contour, and the treatment targets: PTV$_H$ (in purple - receiving the higher dose) and PTV$_L$ (in cyan - receiving the lower dose). The marked urethra and rectum are structures to be spared.

**The iterative simulated annealing algorithm:**

Starting with an initial temperature, $T_0$, the pseudo code for the SA algorithm is as follows:

* Initialize plan seed locations
* While *stopping criterion* is not met
    For k= 1:*max_iterT*
       While *max_number_of_tries* not reached
- Update seed locations
- Calculate the cost
- If update leads to lower cost: accept
- If update leads to higher cost: accept if probability *p>rand*
- Keep track of global best solution: *best_solution*

       If *max_number_of_tries* reached: start from *best_solution*
    Update T

The acceptance probability is $p = \exp(-\Delta E/T(k))$ where $\Delta E$ is the difference between the cost function value at iteration $k$ and $k-1$. $rand$ is a randomly chosen from a uniform distribution function in the interval (0,1). $T_0$, the initial temperature, is set to 3 using the method described in Ben-Ameur [16], such that the acceptance ratio of increasing cost moves is equal to 0.8. The maximum number of iterations per temperature change, *max_iterT*, is set to 3. The maximum number of iterations for unacceptable plan updates, *max_number_of_tries*, is set to 20 and is mostly reached as the algorithm starts to converge to an optimum.

Details of the components in the pseudo code above are as follows.

*Air kerma strength*

To manage better dose coverage and maintain reasonably low seed densities for the asymmetric dose distribution requirements and small boost regions in our application, dual air kerma strength ($S_k$) sources are used. That is, plans contain a combination of seeds with a high $S_k$ of 0.74 U and, a low $S_k$ of 0.301U or 0.354 U for PTVs more than or less than approximately 34 cm$^3$, respectively. The 34 cm$^3$ threshold was based on in-house data for $S_k$/cm$^3$ as a function of PTV volume for our conventional whole-gland treatment plans. The choice of $S_k$ was based on values available by the source provider. A simplified version of the general TG-43 equation is used for dose computation which includes averaged anisotropy effects and assumes a point source [17]. Values for AgX100 Theragenics (Theragenics Corp., Georgia, USA) seeds are used here.

It is worth noting that despite the similar appearance of seeds in a dual activity implant, distinguishing the seeds in post-implant CT-based dose reconstruction is possible using an in-house seed reconstruction software described in [18].

*Initial seed locations*

Certain rules developed through experience are followed at our center to create plans that are robust and adaptable in the operating room. These, which we also apply here, include: 1) needles must pass through the PTV; 2) seed adjacency is not allowed in any of the three directions (to limit hot spots); 3) fewer than three neighboring needles in directly adjacent grid locations are permitted (to limit hot spots and trauma); 4) seeds are to be placed no more than 1 cm superior-inferior direction and no more than 0.5 cm in-plane direction outside the PTV (to limit extracapsular dose); 5) off-grid locations are not allowed as this lowers plan robustness in case of potential changes in the prostate images between the TRUS image acquisition and implant.

With these constraints in mind, initialization of the algorithm is as follows: needles are initially placed on the basis of the prostate base, apex and largest mid-gland contours (in order) starting from the four most anterior and posterior grid locations in the c (or C) and d (or E) columns and moving laterally following the boundary. Grid points closest to the PTV contour are chosen. The needles are then filled from both ends toward the center, starting from the most anterior needles toward the posterior. Considering seed adjacency restrictions, we used this order of needle filling to ensure adequate anterior dose coverage. In practice it was seen that the anterior contour, being generally narrower than contour posterior, is more prone to under dose. Finally, regions with sparse seed density are filled.

The higher $S_k$ is assigned to seeds at the base and apex slice and inside (and within one grid space exterior to) $PTV_H$, and the lower $S_k$ is assigned to the remaining seeds.

Figure 2 shows an example of a plan initialization. Note how needles and seeds are placed closer to the PTV.

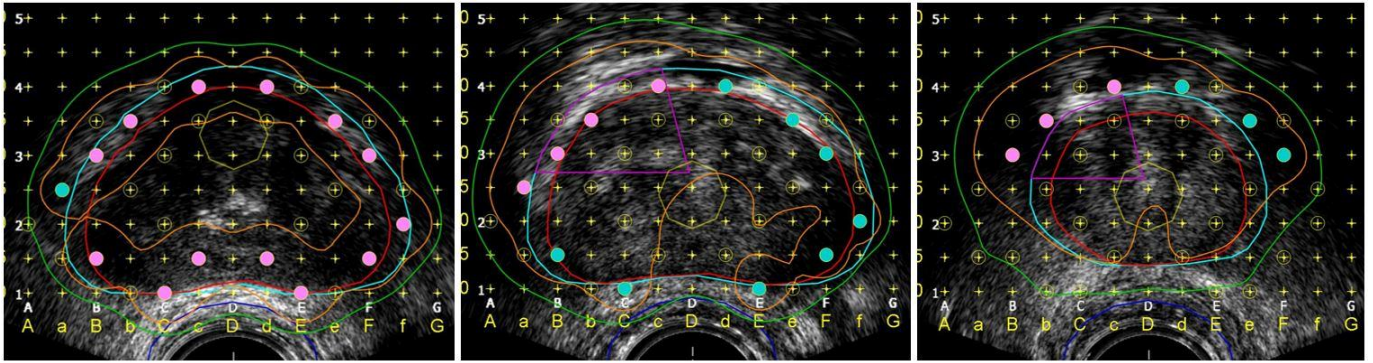

Figure 2: An example of an automatic plan initialization for a AIQ $PTV_H$. Yellow circles indicate needle tracks. Where filled, seeds are located (pink: high $S_k$, cyan: low $S_k$). The $iso_{100}$ and $iso_{150}$ lines are shown in green and orange, respectively. A typical initialization does not aim to optimize the plan, and typically overdoses the gland. The algorithm then optimizes the plan to meet the desired criteria.

*Updating seed locations*

At each iteration, the current plan is modified to derive six candidate plans $P_m, m = 1 \dots 6$ which include: 1&2) randomly select and relocate a seed to an empty grid point (high and low $S_k$); 3) remove the selected seed; 4&5) randomly select an empty grid location and add a seed (high and low $S_k$); 6) randomly select a seed and flip the $S_k$ (i.e. low to high and high to low activity). The candidate leading to the lowest cost is selected.

Single seeded needles are undesirable in LDR-PB planning unnecessarily increasing the needle density and being more prone to post-implant seed displacement. When selecting a seed for plan modification, priority is given to seeds from needles that have remained single seeded the longest.

*Cost function*

The total cost $C$, at iteration $i$ and plan modification $P_m$, is the weighted sum of deviation from the planning constraints, $f_c(i, P_m)$, derived from those used in standard LDR-PB boost treatment planning with modifications for focal treatment (eq. 1). Table 1 shows the $c = 14$ constraints and their nominal values $N$ ($N$ being either the high, low, or equal value $N_h$, $N_l$, $N_e$, respectively) use in computing $f_c$.

$$C(i, P_m) = \sum_{c=1}^{14} w_c\, f_c(i, P_m) \tag{1}$$

Table 1: Criteria used in deriving the cost function

|  | **Constraints applied in the algorithm, $N$ : Nominal values ($N_l, N_h,\ or\ N_e$)** |
|---|---|
| 1 | $n_{ss}/\ n_{T\_ss}$ : Number of single seeded needles (normalized to total possible single seeded needles), $N_h$: 0 |
| 2 | Total $S_k$, $N_h$: < maximum defined based on in-house data for $S_k/cm^3$ as a function of PTV volume |
| 3 | $n_n$: Number of needles (normalized to total possible number of needles), $N_h$: < maximum needles based on in-house data |
| 4 | $n_{neigh}$: Number of neighbors per needle (normalized to total number of neighbors), $N_h$: <3 |
| 5 | $PTV_L\ V_{100}$, $N_l$ : ≥ 97% |
| 6 | $PTV_{H+L}\ g_{100}$: $PTV_L+PTV_H$ iso$_{100}$ geometric adherence, $N_e$: 100% |
| 7 | $PTV_H\ V_{150}$, $N_l$ : ≥ 85% |
| 8 | $PTV_H\ g_{150}$: $PTV_H$ iso$_{150}$ geometric adherence, $N_h$: 100% |
| 9 | $PTV_L\ V_{150}$, $N_h$: < 45% |
| 10 | Urethra $V_{150}$, $N_h$: < 5% |
| 11 | Posterior $PTV_H g_{100}$: Posterior $PTV_{100}$ geometric adherence, $N_e$: 100% |
| 12 | $EI_{150}$: Iso$_{150}$ External index, $N_e$: 0 |
| 13 | $PTV_{-2mm}g_{100}$: PTV minus margin of 2mm iso$_{100}$ geometric adherence, $N_e$: 100% |
| 14 | $n_{so}$: number of seeds outside the PTV (normalized to maximum number of seeds outside the PTV), $N_e$: 0 |

Standard dose metrics recommended by the AAPM [19] (e.g. $V_{100}$ and $V_{150}$ in entries 5, 7, 9, 10 in Table 1) set numerical goals for dose coverage but they do not indicate the shape of coverage. A PTV $V_{100}$ of 95% ensures that 95% of the PTV volume receives at least 100% of the prescribed dose. However, different plans one can deliver this value, some being more acceptable than others. In order to consider the shape of the isodose surface covering the target volumes, we include geometric constraints in the cost function calculation (items 6, 8, 11 and 13 in Table 1). These include:

$$PTV_{H+L}\ g_{100} = \left(\sum_{(x,y)} |\ (iso_{100})_{mask} - (PTV_{H+L})_{mask}\ | / PTV\right)^2 \quad (2)$$

$$PTV_H\ g_{150} = \left(\sum_{(x,y)} |\ (iso_{150})_{mask} - (PTV_H)_{mask}\ | / PTV_H\right)^2 \quad (3)$$

$$Posterior\ PTV_H\ g_{100} = \left(\sum_{(x,y) \in PTV\ posterior} |\ (iso_{100})_{mask} - (PTV\ or\ PTV_{+3mm})_{mask}\ |\right)^2 \quad (4)$$

$$PTV_{-2mm}\ g_{100} = \left(\sum_{(x,y) \in PTV_{-2mm}} |(iso_{100})_{mask} - (PTV_{-2mm})_{mask}| / PTV\right)^2 \quad (5)$$

where ( )$_{mask}$ indicates a binary mask of the contour within the parentheses. $PTV_{+3mm}$ ($PTV_{-2mm}$) is the PTV plus (minus) a margin of 3mm (2mm); $PTV_{H+L}$ is the union of $PTV_H$ plus a margin of 5mm and $PTV_L$ plus a margin of 2mm. These contours are automatically derived.

In Eq. (2) an ideal value of zero ($N_e$:100% adherence) indicates overlap of $iso_{100}$ with $PTV_{H+L}$. $iso_{100}$ is the contour/surface on which 100% of the nominal prescribed dose is delivered, e.g. the green line in Figure 2.

Eq. (3) encourages 150% of the nominal prescribed dose only to $PTV_H$. In Eq. (4) $iso_{100}$ is encouraged to follow the posterior boundary of the PTV in the mid-gland slices and for superior/inferior slices where the PTV is farther from the rectum, the $PTV_{+3mm}$ contour.

Eq. (5) aims for complete coverage of the $PTV_{-2mm}$ by the $iso_{100}$. This is used to avoid significant dipping of the $iso_{100}$ and allow 'skimming' (by 2mm) of the PTV by the $iso_{100}$ which both can lead to the same $V_{100}$, while the latter is acceptable and the former is not.

Items 12 and 14 in Table 1 encourage a more compact plan by measuring the extra-prostatic dose [20] and limiting the number of seeds outside the PTV.

All dose metrics used in the cost function are normalized as needed, either to the nominal values or maximum possible values. Weights, $w_i$, are chosen by experimentation such that the cost component values fall approximately within the same range but also to reflect the importance each of the criteria.

*Stopping criteria*

The algorithm stops if one of these conditions is met: 1) temperature T reaches below a threshold (0.002); 2) standard deviation of the cost in the last 40 iterations is less than 0.02; 3) the algorithm restarts from the best found solution 3 times for the same temperature value.

The output of the algorithm is the best found solution.

**Evaluation**

Evaluation metrics are as shown in Table 2. Standard dose metrics were modified for focal treatment: $PTV_L$ $V_{100} \geq 0.97$ and $PTV_H$ $V_{150} \geq 0.85$, $PTV_L$ $V_{150} < 0.45$. Programming was done with Matlab 2018b on a standard PC (Intel® Core™ i7 CPU, 8 GB RAM).

*Single activity needle loading*

The described planning approach does not restrict the air kerma strength within the needles (we will refer to this as "dual $S_k$ needle" plans). Restricting each needle to contain seeds of the same, either of the high or low, $S_k$ (i.e. "single $S_k$ needle" plans) simplifies needle loading in practice. We compare these two needle loading plan types.

*Plan initialization*

To assess whether user intervention would be beneficial or a less sophisticated planning approach would be sufficient, the proposed automatic plan initialization was compared to two other initialization techniques on 20 consecutive cases: 1) manual initialization, where the standard approach of placing needles in the PTV periphery and filling sparse central areas used in the algorithm was manually applied (with no additional manual optimization), and 2) Seattle-based initialization, where needles were placed simply at alternating grid points within the PTV (also with no additional manual optimization).

*Plan rating*

The 20 consecutive cases planned using the three different initialization methods were rated by two radiation oncologist blind to the initialization technique and each other's rankings. Plans were ranked as unacceptable, acceptable with modifications, acceptable without modifications.

**RESULTS**

Figure 3 shows two optimized plans using automatic initialization (top row corresponding to the case initialized in Figure 2).

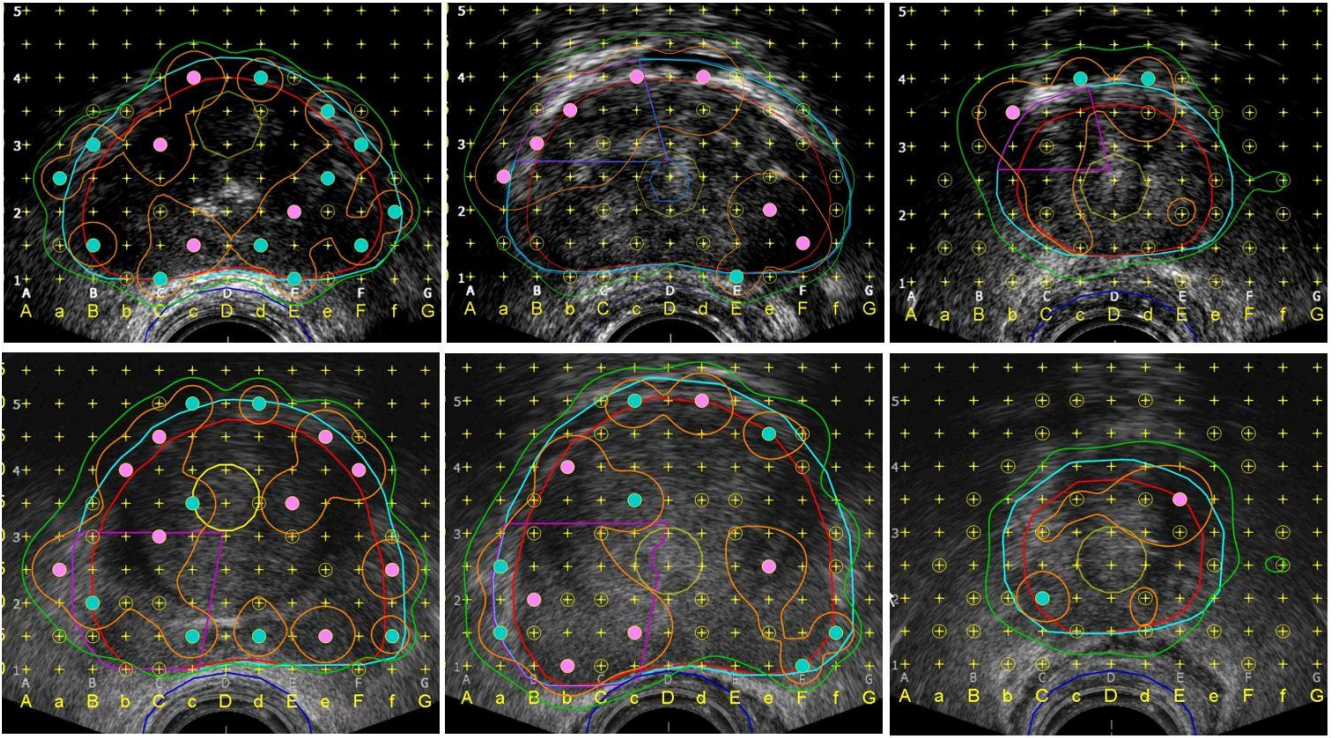

Figure 3: Automatic plans generated for two cases (AIQ and PSQ PTV$_H$) shown at the base (left), mid-gland (middle) and apex (right). Where filled, seeds are located (pink: high $S_k$, cyan: low $S_k$). The iso$_{100}$ and iso$_{150}$ lines are shown in green and orange, respectively. The top plan (34cm$^3$ prostate) consists of 27 needles and 69 sources (25 high activity) and the bottom plan (51cm$^3$ prostate) consists of 32 needles and 80 sources (36 high activity). 0.740 U and 0.301 U sources were used for both.

Table 2 shows automatic plan characteristics for all quadrants of 46 cases (184 plans) and stratified for each of the quadrants. Nominal values for PTV$_L$ V$_{100}$ and V$_{150}$, PTV$_H$ V$_{150}$ and urethra V$_{150}$ are those listed in Table 1. Results are shown for both dual $S_k$ and single $S_k$ needles. No clinically significant difference was observed between the two needle loading plan types or within the quadrants.

In both needle loading plan types the rectal dose is slightly higher for the two posterior quadrants. In our standard practice, the rectum V$_{50}$ is typically below 10%, however with a large variation. This is mostly due to variation in the distance between the rectum and prostate posterior related to the amount of upward TRUS probe pressure applied to obtain acceptable imaging quality.

Total $S_k$ is within permitted values in both needle loading plan types. While the optimization time, number of needles and percent cases with single seeded needles is slightly less for dual $S_k$ needle plans compared to single $S_k$ needle plans, this difference is not clinically significant.

Table 2: Characteristics (95% CI) of automatic plans for all quadrants and stratified for each of the quadrants. Single $S_k$ needle plans and dual $S_k$ needle plans are compared.

| Plan characteristics (nominal values) \ Region | PTV$_L$ V$_{100}$ (≥0.97) | PTV$_H$ V$_{150}$ (≥0.85) | PTV$_L$ V$_{150}$ (<0.45) | Urethra V$_{150}$ (<0.05) | Rectum V$_{50}$ | # seeds, # needles, % cases with more than max # needles | Duration [min] | % cases with single seeded needles, max # single seeded needles |
|---|---|---|---|---|---|---|---|---|
| **Dual $S_k$ needle loading** | | | | | | | | |
| All quadrants | 0.98±0.00 | 0.86±0.09 | 0.41±0.01 | 0.06±0.00 | 0.18±0.01 | 62.5±2.1, 25.8±0.7, 2.7% | 10.6±0.6 | 34.2, 3 |
| AIQ | 0.98±0.00 | 0.88±0.01 | 0.42±0.01 | 0.05±0.01 | 0.16±0.02 | 62.2±4.1, 25.5±1.4, 0% | 8.9±1.0 | 34.8, 2 |
| ASQ | 0.98±0.00 | 0.88±0.01 | 0.40±0.01 | 0.06±0.00 | 0.16±0.02 | 62.2±4.2, 25.9±1.5, 6.5% | 11.7±1.6 | 39.1, 2 |
| PIQ | 0.98±0.00 | 0.85±0.00 | 0.42±0.01 | 0.07±0.01 | 0.20±0.02 | 62.6±4.1, 25.9±1.4, 0% | 11.7±1.2 | 34.8, 3 |
| PSQ | 0.99±0.00 | 0.85±0.00 | 0.41±0.02 | 0.06±0.00 | 0.19±0.02 | 63.0±4.1, 25.9±1.4, 4.3% | 10.2±0.9 | 28.3, 2 |
| **Single $S_k$ needle loading** | | | | | | | | |
| All quadrants | 0.97±0.00 | 0.86±0.00 | 0.42±0.01 | 0.07±0.00 | 0.18±0.01 | 68.0±2.3, 26.9±0.7, 14.1% | 11.7±0.6 | 40.2, 4 |
| AIQ | 0.98±0.00 | 0.87±0.01 | 0.43±0.01 | 0.07±0.01 | 0.15±0.02 | 67.2±4.7, 26.6±1.4, 13.0% | 11.4±1.3 | 34.8, 2 |
| ASQ | 0.97±0.00 | 0.88±0.01 | 0.40±0.01 | 0.06±0.01 | 0.15±0.02 | 68.7±4.8, 26.9±0.1, 17.4% | 11.8±1.2 | 23.9, 2 |
| PIQ | 0.98±0.00 | 0.85±0.01 | 0.42±0.01 | 0.07±0.01 | 0.21±0.00 | 67.7±4.6, 26.8±1.3, 6.5% | 11.8±1.3 | 50.0, 4 |
| PSQ | 0.97±0.00 | 0.85±0.00 | 0.41±0.02 | 0.07±0.01 | 0.22±0.02 | 68.5±4.7, 27.1±1.3, 19.6% | 11.8±1.3 | 52.2, 4 |

Table A.1 in the Appendix shows plan characteristics for the first 20 cases (plan type dual $S_k$ needle) comparing three different initialization schemes. Automatic initialization leads to plans with slightly more seeds. Manual initialization leads to hotter $PTV_L$ $V_{150}$ but cooler rectum. Optimization time is similar among all initialization schemes; however, manual initialization requires at least extra 5 minutes.

Based on the radiation oncologists' ratings of the three different initialization schemes, regardless of initialization method, out of the 80 plans (20 cases four quadrants each), plan acceptability was as follows for RadOnc 1/RadOnc 2 (with more experience), respectively: 92% / 84% of the automatic plans were acceptable ( 43% / 50% requiring no modifications) and 8% / 16% were unacceptable. Manual initialization led to the least unacceptable plans (3 and 4 plans for each radiation oncologist, respectively) and the most acceptable without modification (47 and 48 plans). Acceptability of automatic initialization was in general better than Seattle. See Table A.2 for details.

Details per quadrant boost are shown in Figure 4. Based on the radiation oncologists' reviews, modifications were mostly minor being either: 1) anterior displacement of one needle to reduce rectal dose; 2) removal of a seed placed too inferior in the apex.

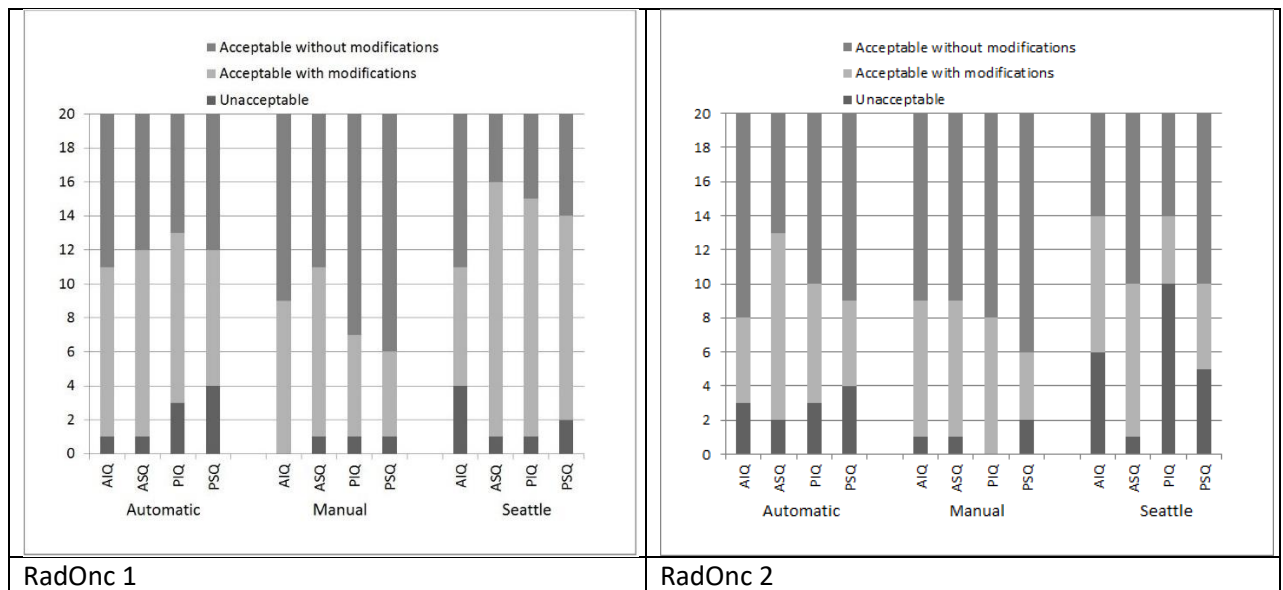

Figure 4: Radiation oncologists' ratings of plans created using three initialization schemes: Automatic, Manual, Seattle.

Characteristics of the automatic plans stratified by prostate volume are shown in Table 3. Results over all cases and quadrants are repeated from Table 2 for comparison. The lower rectal dose for smaller prostates is most likely due to the larger distance between the rectum and prostate. The dose to the urethra and $PTV_L$ and the number of single seeded needles are higher in smaller prostates, as expected. Achieving a balance between seed density, coverage and limiting dose to the organs at risk is more difficult in smaller volumes with fewer possible seed locations.

Table 3: Automatic plan characteristics stratified for prostate volume (automatic initialization scheme and dual Sk needle loading).

| Plan characteristics (nominal values) \ Volume | $PTV_L\ V_{100}$ ($\geq 0.97$) | $PTV_H\ V_{150}$ ($\geq 0.85$) | $PTV_L\ V_{150}$ ($<0.45$) | Urethra $V_{150}$ ($<0.05$) | Rectum $V_{50}$ | # seeds, # needles, % cases with more than max # needles | Duration [min] | % cases with single seeded needle, max # single seeded needles |
|---|---|---|---|---|---|---|---|---|
| All quadrants | 0.98±0.00 | 0.86±0.09 | 0.41±0.01 | 0.06±0.00 | 0.18±0.01 | 62.5±2.1<br>25.8±0.7<br>2.7% | 10.6±0.6 | 34.2 (3) |
| vol≥40 [cm$^3$] (n=64) | 0.98±0.00 | 0.86±0.01 | 0.41±0.01 | 0.05±0.00 | 0.21±0.02 | 76.3±1.3<br>30.5±0.4<br>4.7 | 13.2±1.1 | 17.2 (1) |
| 25≤vol<40 [cm$^3$] (n=76) | 0.98±0.00 | 0.87±0.01 | 0.41±0.01 | 0.06±0.00 | 0.18±0.01 | 62.7±1.5<br>25.9±0.5<br>2.6% | 10.5±0.6 | 39.5 (2) |
| 25<vol [cm$^3$] (n=44) | 0.98±0.00 | 0.86±0.01 | 0.43±0.02 | 0.07±0.01 | 0.13±0.02 | 42.2±1.9<br>18.9±0.9<br>0 | 7.1±1.0 | 50 (3) |

**Discussion**

The proposed method is based on our centre's successful whole-gland LDR-PB planning approach [21]. Standard planning constraints were modified based on our previous experience with focal planning [4,5], particularly regarding challenges with the small focal boost target. For example, our standard for whole-gland LDR-PB boost is $60 < V_{150} < 70\%$. However, for focal planning, we aim for conformal coverage of a small sub-region of the PTV with the 150% isodose, while keeping the remainder of the PTV and adjacent organs at risk below 150%; hence the expected $PTV_H\ V_{150}$ was set to 85%.

Smaller prostates with a correspondingly smaller $PTV_H$ were more difficult to automatically plan. Either target coverage or the dose to organs at risk had to be sacrificed (Table 3). It must be noted, however, that the data set used here included a higher proportion of extreme prostate shapes and sizes. We expect better overall results in practice with more typical prostate shapes and volumes.

Planning is highly dependent on contouring. In practice, symmetric positioning of the prostate above the ultrasound probe and creation of symmetric contours and plans is performed to reduce the effect of contouring variabilities. This is not the case for focal LDR-PB with asymmetry in the treatment target and the required dose distribution. Hence, one needs to consider the potential effect of contouring on the dose delivered to various structures. For example, based on the radiation oncologists' ratings, most required plan modifications were to further reduce the rectal dose, particularly at the apex, due to rectal proximity to the posterior prostate contour. A solution, sometimes used in practice, is to assume that a larger separation between the posterior prostate and rectum is achievable at the time of implant (due to the intra-operative relaxed state of the patient as opposed to the tense awake state at imaging) and shift the plan template (and plan) anteriorly during planning.

A challenge in obtaining a desirable plan is that while guidelines exist for standard treatment, planning preferences are inconsistent among different centres or physicians. In our standard practice physicians are offered a few plan variations to choose from which they may modify before and during implantation. An optimization algorithm, at the best, can produce plans that satisfy given constraints and minor modifications may be needed since not all physician preferences can be mathematically modeled. One potential source of ranking bias is that the radiation oncologists, despite being blinded, may have developed a preference for certain features used in our manual planning algorithms at our institution. Furthermore, plans are being ranked based on the radiation oncologists' experience on standard LDR-PB and may change after more experience with focal LDR-PB boost plans.

Looking from a different perspective, tuning the cost component weights can provide the flexibility of customizing plan creation. Plan variations can be offered based on individual or institution specific planning preferences. Novel machine learning methods can also be used for extracting features of desirable plans and even further, centre specific or physician specific planning trends. Such approaches generally require a large training set which, given that focal LDR-PB is not yet standard of care, is not yet available to us.

Most commercial needle loading facilities provide needles loaded with the same air kerma strength seeds. Our results show that plans with single $S_k$ needles are slightly – though not significantly- inferior in terms of coverage, duration, maximum number of needles and number of single seeded needles. This is expected as a dual $S_k$ needle planning has one more degree of freedom. For focal targets, the use of dual $S_k$ plans in either of the two forms is the feasible approach to avoid high seed density while achieving good coverage, as we showed in our previous studies [4,5]. A single $S_k$ needle loading approach may be more feasible in practice in terms of needle preparation.

Our choice of seed strengths was based on our prior planning experience which allowed us to use known initialization strategies. Future work could include optimization of source strengths.

The proposed approach creates plans within the order of 10 minutes. While this is acceptable for pre-operative planning, shorter planning times are preferred for intra-operative planning. A trivial solution is the use of a higher performing system. However, while fast -treatment planning techniques are desirable for intra-operative planning, one must be aware of the additional operating room time required for needle loading, whereas in the case of pre-operative planning, this is not an issue.

The literature on automatic focal LDR-PB planning is very limited. Betts et al. [13] propose the use of a radiobiological model within an iterative optimization method. They assess the robustness of their plans to post-operative seed displacements. However, they do not apply geometric constraints on the rectum or urethra, which can be particularly important for tumors adjacent to these structures and furthermore, single strength seeds are used.

Currently, our method is particularly designed for the focal LDR-PB boost; however, a similar approach can be applied to whole gland treatment or focal LDR-PB for low-risk PCa (where only the diseased region is targeted and the rest of the gland is spared).

**Conclusions**

In this article we developed a simulated annealing-based method for automatic focal LDR-PB boost planning. Plans were optimized for both dosimetric constraints and geometric constraints leading to a high rate of acceptability (average 88% over two radiation oncologists) with very minor modifications (average 46% not requiring modification). While manual initialization improves acceptability (91%), fully automatic plans are also mostly acceptable (74%). The use of dual $S_k$ seed plans is technically feasible, whether in the form of dual or single $S_k$ needle loading, to achieve appropriate dose coverage in such asymmetric targets with focal boost regions of small volume, but requires further exploration in terms of clinical implementation.

**Disclosure**

The authors have no relevant conflicts of interest to disclose. This study was supported by the Canadian Institutes of Health Research and BC Cancer Foundation.

## Appendix

Table A.1: Plan characteristics for the first 20 cases comparing three different initialization schemes used for automatic planning. Statistically significant differences from the automatic initialization are indicated in bold (paired t-test, p-value<0.05).

| Plan characteristics (nominal values) / Initialization method | $PTV_L$ $V_{100}$ ($\geq$0.97) | $PTV_H$ $V_{150}$ ($\geq$0.85) | $PTV_L$ $V_{150}$ (<0.45) | Urethra $V_{150}$ (<0.05) | Rectum $V_{50}$ | # seeds, # needles, % cases with more than max # needles | Duration [min] | % cases with single seeded needle, max # single seeded needles |
|---|---|---|---|---|---|---|---|---|
| 1 – Automatic | 0.98±0.00 | 0.87±0.01 | 0.42±0.01 | 0.06±0.00 | 0.18±0.02 | 62.2±3.7 / 25.5±1.2 / 1.25% | 11.0±1.1 | 25.0, 3 |
| 2 – Manual | 0.98±0.00 | 0.87±0.01 | **0.44±0.01** | 0.05±0.00 | **0.17±0.01** | **60.7±3.5 / 25.0±1.2** / 0 | 11.7±1.0 + ~**5mins manual initialization** | 35.0, 2 |
| 3 - Seattle | 0.98±0.00 | **0.86±0.01** | 0.43±0.01 | **0.08±0.01** | 0.18±0.01 | **60.8±3.4** / 25.6±1.2 / 6.25% | 11.5±1.0 | 16.3, 3 |

Table A.2: Plan ranking results for two radiation oncologists RadOnc1 and RadOnc2. A total of 20 plans, four quadrants each, were initialized in three ways: automatic, manual and the Seattle method. Plans were ranked as unacceptable, acceptable with modifications and acceptable without modifications.

|  | RadOnc 1 | | | RadOnc 2 | | |
|---|---|---|---|---|---|---|
| **Initialization method** | **Automatic** | **Manual** | **Seattle** | **Automatic** | **Manual** | **Seattle** |
| **# unacceptable** | 9 | 3 | 8 | 12 | 4 | 22 |
| **# acceptable with modification** | 39 | 30 | 48 | 28 | 28 | 26 |
| **# acceptable without modification** | 32 | 47 | 24 | 40 | 48 | 32 |